\documentclass[lettersize,journal]{IEEEtran}
\usepackage{amsmath,amsfonts}
\usepackage{algorithmic}
\usepackage{array}
\usepackage[caption=false,font=normalsize,labelfont=sf,textfont=sf]{subfig}
\usepackage{textcomp}
\usepackage{stfloats}
\usepackage{url}
\usepackage{verbatim}
\usepackage{graphicx}

\usepackage{amssymb}
\usepackage{multicol}
\allowdisplaybreaks
\def\BibTeX{{\rm B\kern-.05em{\sc i\kern-.025em b}\kern-.08em
		T\kern-.1667em\lower.7ex\hbox{E}\kern-.125emX}}
\usepackage{balance}
	
\begin{document}
	\title{Randomized Power Transmission with Optimized Level Selection Probabilities in Uncoordinated Uplink NOMA 
		}
	\author{Noura Sellami,  LETI Lab., ENIS, University of Sfax, Tunisia, noura.sellami@enis.tn\\
		Mohamed Siala, MEDIATRON Lab., SUP'COM, University of Carthage, Tunisia, mohamed.siala@supcom.tn}	
\maketitle			
	\begin{abstract}
		We consider uncoordinated random uplink non-orthogonal multiple access (NOMA) systems using a set of predetermined power levels. We propose to optimize the probabilities of selection of power levels in order to minimize performance metrics as block error probability (BLEP) or bit error probability (BEP). When the multiuser detection algorithm at the BS treats at most two colliding users' packets, our optimization problem is a quadratic programming problem. For more colliding users' packets, we solve the problem iteratively. Our solution is original because it applies to any multiuser detection algorithm and any set of power levels.
			\end{abstract}
	\section{Introduction}
	Wireless NOMA has been extensively studied since it increases spectral efficiency by allowing several users to share the same time and frequency resources \cite{Liu2017}\cite{ChoiISWCS2017}. It also facilitates massive connectivity in machine-type communications (MTC) for Internet of Things (IoT) systems \cite{Mostafa2019}. Since NOMA uses the power domain to achieve multiple access, multiuser detection techniques, as successive interference cancellation (SIC) or joint detecting (JD), are used at the receiver, to recover the packets of active users \cite{Tegos2020}. 
In most existing works, coordinated transmissions with known channel state information (CSI) have been considered. However, uncoordinated random access schemes are more suitable for MTC due to low signaling overhead when devices have short packets to transmit \cite{Choi2017}. Moreover, MTC has large number of devices but their activity is sparse \cite{Wunder2015}, thus random access is generally used for uplink transmissions. In \cite{Choi2017}, an uplink NOMA based random access scheme with multichannel ALOHA has been proposed where each active user chooses one power level among predetermined power levels uniformly at random. 
 The problem of optimizing the probabilities of choosing power levels by active users has been studied in \cite{Mai2021}. However, the proposed solution can only be applied when perfect SIC is used at the base station (BS), which is not reasonable in practical cases where SIC is imperfect or JD can be preferred \cite{Tegos2020,Semira2021}.
 
	In this letter, we consider an uplink NOMA random access system involving multiple users and multiple power levels. Our aim is to optimize the set of probabilities of selection of power levels such that performance metrics as BLEP or BEP are minimized. The originality of our solution stems from the fact that it can be applied for any detection algorithm used at the BS and for any choice of the set of power levels. We show that in the case where the multiuser detection algorithm at the BS handles two colliding users' packets, the optimization problem is a quadratic programming problem. When more colliding users' packets can be treated, we propose to solve the optimization problem iteratively.
	
	The remainder of this letter is organized as follows. In section 2, we introduce
	the system model. In section 3,  we present the optimization problem. In section 4, we give a simple case study to assess the performance of our solution. In section 5, we provide simulations results before concluding this letter in section 6. Throughout this letter, vectors are underlined and matrices are bold, lower or upper case. Moreover, $(.)^*$, $(.)^T$ and $(.)^H$ denote respectively conjugation, transposition and trans-conjugation.
	
		\section{System model}
	We consider a wireless uplink NOMA random access system consisting of a BS and $N$ users. The received signal at the BS, in a slot, is given by 
	\begin{equation}
	x=\sum_{n\in \tau}h_n\sqrt{p_n}s_n+w,
	\end{equation}
	where $\tau$ is the set of active users, $s_n$ is the signal transmitted by user $n$, $p_n$ is its transmit power, $h_n$ is the channel tap from user $n$ to the BS and $w$ is the additive white Gaussian noise whose spectral density is ${N_0}$.   
	We consider the time division duplexing (TDD) mode and assume that each user $n$ perfectly knows the channel $h_n$. As in \cite{Choi2017}, there are $Q$ predetermined power levels, received at the BS, that are denoted by 
	\begin{equation}
	v_1>...>v_Q>0.
	\end{equation}
	We assume that each active user $n$ can randomly choose one of the power levels $v_q$, for $1 \leq q \leq Q$, for random access. The transmission power, when the selected level is $v_q$, is given by
		\begin{equation}
p_n=\frac{v_q}{\alpha_n},
	\end{equation}
	where $\alpha_n=|h_n|^2$. Thus, the received power at the BS is equal to $v_q$. Let $P_{q}$ be the probability that an active user chooses the power level $v_q$, for $1 \leq q \leq Q$, and $\underline{P}=(P_{1},\ldots,P_{Q})^T$. In the next section, we propose to find $\underline{P}$ that minimizes performance metrics as average BLEP or average BEP. Without loss of generality, we focus in the following on the minimization of the average BLEP.

	\section{Optimization problem}

Our aim is to find the optimal set of probabilities that minimizes the overall average BLEP $\overline{P}_E$. We assume that at least one user is active. The number $k$ of other active users can be approximated, for a large $N$, by a Poisson distribution with parameter $\lambda$ as follows
\begin{equation}
 p_{\lambda}(k)=\frac{\mathrm{e}^{-\lambda}\lambda^k}{k!}.
\end{equation}

Then, the overall average BLEP is given by

\begin{equation}
\overline{P}_E=\sum_{k=0}^{\infty}  \frac{\mathrm{e}^{-\lambda}\lambda^k}{k!} \overline{P}_E^{(k)},
\label{PE}
\end{equation}
where $\overline{P}_E^{(k)}$ is the average BLEP obtained when $k+1$ users, for example users $0,1,\cdots,k$, are transmitting simultaneously their packets ($k$ collisions).
It is given by
\begin{equation}
\overline{P}_E^{(k)}=\sum_{i_0,i_1,\cdots,i_{k}=1}^{Q} P_{i_0}P_{i_1}\cdots P_{i_k} P_E\left( \gamma_{i_0},\gamma_{i_1}, \cdots ,\gamma_{i_{k}}\right), 
\label{eqPEk}
\end{equation}
where $P_E\left( \gamma_{i_0},\gamma_{i_1}, \cdots ,\gamma_{i_{k}}\right)$ is the BLEP when $k+1$ users are transmitting simultaneously and user $k'$, for $0 \leq k'\leq k$, chooses power level $v_{i_{k'}}$, with $1\leq i_{k'} \leq Q$ and $\gamma_{i_{k'}}=\frac{v_{i_{k'}}}{N_0}$ are the received signal to noise ratio (SNR) levels at the BS. Notice that $P_E\left( \gamma_{i_0},\gamma_{i_1}, \cdots ,\gamma_{i_{k}}\right)$ depends on the detection algorithm used at the BS. 

Our aim is to minimize $\overline{P}_E$ subject to the following constraints:
\begin{subequations}
	\label{opt1}
	\begin{align}
	\underset{\underline{P}}{\text{Minimize}} \quad & \overline{P}_E \\
	\text{subject to} \quad & C_1:  \sum_{q=1}^{Q} P_{q}=1 \\
	&C_2:  \sum_{q=1}^{Q} P_{q}\gamma_{q}= \overline{\gamma} \\
	&C_3:  P_{q} \geq 0, \quad q = 1,2,\dots,Q.
	\end{align}
\end{subequations}

In the minimization problem, $C_1$ indicates that the sum of the probabilities is equal to one, $C_2$ implies that the average SNR at the receiver is equal to a target SNR $\overline{\gamma}$ and $C_3$ ensures that the probability of choosing a power level is a non-negative value.

We assume in the following that the multiuser detection algorithm at the BS handles a set number $K+1$ of packets in collision and $\overline{P}_E^{(k)}=1$, for $k\geq K+1$, is then independent of $\underline{P}$. We consider the truncated expression of (\ref{PE})

\begin{equation}
\overline{P}_{E_{T}}=\sum_{k=0}^{K}  \frac{\mathrm{e}^{-\lambda}\lambda^k}{k!} \overline{P}_{E}^{(k)}.
\label{PET}
\end{equation}
In the following, we present efficient solutions to solve the optimization problem according to the number of colliding users packets handled by the multiuser detector at the BS.

\subsection{Collision of at most two  users' packets ($K \leq 1$)}
 We first assume that the multiuser detector algorithm can treat at most two users' packets in collision. Then, the overall average truncated BLEP is given by

\begin{equation}
\overline{P}_{E_{T}}=\sum_{k=0}^{1}  \frac{\lambda^k}{k!} \mathrm{e}^{-\lambda} \overline{P}_E^{(k)},
\label{PET_K1}
\end{equation}
where $\overline{P}_E^{(0)}=\underline{P}^T \underline{M_0}$ with $\underline{M_0}=\left( P_E(\gamma_{1}),\ldots,P_E(\gamma_{Q})\right) ^T$
		and
	\begin{equation}
	\overline{P}_E^{(1)}=\sum_{i_0,i_1=1}^{Q} P_{i_0}P_{i_1} P_E\left( \gamma_{i_0},\gamma_{i_1}\right).  
	\end{equation}
	Let $\underline{\gamma}=(\gamma_{1},\ldots,\gamma_{Q})^T$, $\mathbf{M_1}$ be the $Q\times Q$ matrix  whose entry in row $i_0$ and column $i_1$ is $P_E\left( \gamma_{i_0},\gamma_{i_1}\right) $ and $\underline{1}_Q$ denote the all-ones column vector of dimensions $Q\times 1$.
	The minimization problem can be rewritten as 

	\begin{subequations}
		\label{opt2}
			\begin{align}
		\underset{\underline{P}}{\text{Minimize}} \quad & \mathrm{e}^{-\lambda}\underline{P}^T\underline{M_0}+\lambda\mathrm{e}^{-\lambda}\underline{P}^T\mathbf{M_1}\underline{P} \\
		\text{subject to} \quad & \underline{P}^T \underline{1}_Q=1 \\
		& \underline{P}^T \underline{\gamma}=\overline{\gamma} \\
		& P_{q}\geq 0, \quad q = 1,2,\dots,Q.
		\end{align}
	\end{subequations}
	This problem is a quadratic programming problem. We can solve
	it to obtain its globally optimal solution efficiently
	by using numerically stable optimization methods \cite{Boyd2004}. In our
	simulations in section V, we use \texttt{quadprog} in MATLAB to solve
	the problem.
\subsection{Collision of more than three  users' packets ($K \geq 2$) }
 We first assume that $K=2$. The overall average truncated BLEP is given by

\begin{equation}
\overline{P}_{E_{T}}=\sum_{k=0}^{2}  \frac{\lambda^k}{k!} \mathrm{e}^{-\lambda} \overline{P}_E^{(k)},
\end{equation}
where
\begin{equation}
\overline{P}_E^{(2)}=\sum_{i_0=1, i_1=1,\cdot i_2=1}^{Q}  P_{i_0}P_{i_1} P_{i_2}P_E\left( \gamma_{i_0},\gamma_{i_1},\gamma_{i_{2}}\right).
\label{eqPE3}
\end{equation}

In this case, we propose to solve the problem iteratively. 
\begin{itemize}
	\item iteration 0: find $\underline{P}^{(0)}$ that minimizes $\mathrm{e}^{-\lambda}\underline{P}^T\underline{M_0}+\lambda\mathrm{e}^{-\lambda}\underline{P}^T\mathbf{M_1}\underline{P}$ subject to $C_1$, $C_2$ and $C_3$.\\
	\item iteration $l$, $l\geq 1$: let $\underline{P}^{(l-1)}=\left(P_{1}^{(l-1)},P_{2}^{(l-1)},\cdots,P_{Q}^{(l-1)}\right)^T$ be the optimized value of the vector $\underline{P}$ obtained at iteration $l-1$ and $\overline{P}_E^{(2,l)}$ the expression obtained by replacing in equation (\ref{eqPE3}) $P_{i_0}$, for $1 \leq i_0 \leq Q$, by $P_{i_0}^{(l-1)}$. We can write 
	\begin{equation}
\overline{P}_E^{(2,l)}=\underline{P}^T\mathbf{M_2^{(l)}}\underline{P},
\end{equation}
where $\mathbf{M_2^{(l)}}$ is the $Q\times Q$ matrix  whose entry in row $i_1$ and column $i_2$ is $\sum_{i_0=1}^{Q}  P_{i_0}^{(l-1)}P_E(\gamma_{i_0},\gamma_{i_1},\gamma_{i_2}) $.
	In iteration $l$, we find $\underline{P}^{(l)}$ that minimizes $\mathrm{e}^{-\lambda}\underline{P}^T\underline{M_0}+\lambda\mathrm{e}^{-\lambda}\underline{P}^T\mathbf{M_1}\underline{P}+ \frac{\mathrm{e}^{-\lambda}\lambda^2}{2} \overline{P}_E^{(2,l)}$ subject to $C_1$, $C_2$ and $C_3$. The resulting optimization problem is a quadratic programming problem that can be efficiently solved.
	
\end{itemize}
We can easily extend the iterative algorithm to the case where $K \geq 3$. The first iteration does not change. At iteration $l$, for $l \geq 1$, we replace in the expression of $\overline{P}_E^{(k)}$ given in (\ref{eqPEk}), for $0 \leq k \leq K-2$, $P_{i_k}$, by $P_{i_k}^{(l-1)}$. The obtained optimization problem is then a quadratic programming problem.\\

\textit{Remark:} In practical cases, terminals can have constraints on their transmitted powers $p_n$ such as $p_{min} \leq p_n \leq p_{max}$. Users who are close to the BS tend to transmit at low power but cannot go below the minimum power $p_{min}$. On the other hand, users who are far from the BS tend to transmit at high power but cannot exceed the maximum power $p_{max}$. Users subject to these constraints will not be allowed to choose a certain number of transmit power levels. In this case, one alternative is that each one of these users optimizes the probability distribution of power levels that it is allowed to use by assuming that all other users have the right to use all power levels and the corresponding optimized probabilities vector obtained by solving the optimization problem (\ref{opt1}). Thus, for users subject to these constraints, the optimization problem becomes a linear programming problem. Another alternative is to consider the optimized vector (solution of (\ref{opt1})) and redistribute the probabilities initially assigned to the excluded power levels among the remaining power levels that the terminal can use.

The probabilities $P_E\left( \gamma_{i_0},\gamma_{i_1}, \cdots \gamma_{i_{k}}\right)$ depend on the SNRs and on the multiuser detection algorithm used at the BS. In the next section, we consider a simple case study where binary phase shift keying (BPSK) modulation is used in order to  give simple analytical expressions of these probabilities and validate our proposed solution. In more complicated cases, analytical approximations can be used as well as simulations. 
	\section{Case study}

	We consider the simple case where the used modulation is the BPSK and the channel gains have the same phase. We consider in the following the case study where up to 2 users are transmitting simultaneously. We consider the optimal JD as multiuser detection algorithm. We start by giving expressions of the BEP.

In the case where only one user, say user 0, is transmitting, we denote by	$\gamma_{i_0}$, for $1\leq i_0 \leq Q$, the received SNR at the BS. The BEP is then given by 
\begin{equation}
P_e(\gamma_{i_0})=\frac{1}{2} \left( \mathrm{erfc}\left( \sqrt{\gamma_{i_0}} \right) \right),
\end{equation}
where $\mathrm{erfc}(x)=\frac{2}{\sqrt{\pi}} \int_{x}^{\infty} \mathrm{exp}^{-t^2} dt$ refers to the complementary error function.

In the case where two users, say user 0 and user 1, are transmitting simultaneously their packets, the received powers at the BS are respectively $v_{i_0}$ and $v_{i_1}$, with $1\leq i_0 \leq Q$ and $1\leq i_1 \leq Q$. Their SNRs are respectively $\gamma_{i_0}$ and $\gamma_{i_1}$. Let $P_e(\gamma_{i_0},\gamma_{i_1})$ be the BEP obtained when the optimal JD is used. We consider in the following three cases based on the relative position of $\gamma_{i_0}$ to $\gamma_{i_1}$ and derive the analytical expression of $P_e(\gamma_{i_0},\gamma_{i_1})$ in each case. Figure 1 shows the received constellation (for $w=0$) and the decision boundaries. We assume that the user of interest is user 0. We also assume that all symbols are equiprobable. 

	\begin{itemize}
		\item Case 1: We assume that $\gamma_{i_0}>\gamma_{i_1}$.
		Without loss of generality, we consider the case where $s_0=-1$. As shown in Figure 1.a, the erroneous decision on $s_0$ occurs when the noise causes the received signal to be positive. We can write
		\begin{align}
		p( \hat{s}_0=1, s_0=-1) &= \frac{1}{4}\left(p(\hat{s}_0=1 \mid s_0=-1, s_1=1) \right. \notag\\
		&\quad \left. + p(\hat{s}_0=1 \mid s_0=-1, s_1=-1)\right) \notag\\
		&= \frac{1}{4} \left(p\left(w > \sqrt{v_{i_0}} + \sqrt{v_{i_1}}\right) \right. \notag\\
		&\quad \left. + p\left(w > \sqrt{v_{i_0}} - \sqrt{v_{i_1}}\right) \right).
		\end{align}
		Thus,
		\begin{align}
            P_e(\gamma_{i_0},\gamma_{i_1}) &= \frac{1}{4} \Big( \mathrm{erfc}(\sqrt{\gamma_{i_0}} + \sqrt{\gamma_{i_1}}) \nonumber \\
            &\quad + \mathrm{erfc}(\sqrt{\gamma_{i_0}} - \sqrt{\gamma_{i_1}}) \Big).
        \end{align}		
		\item Case 2: We assume that $\gamma_{i_0}<\gamma_{i_1}$. Based on Figure 1.b, we obtain
\begin{equation}
\begin{split}
p( \hat{s}_0=1, s_0=-1)= \frac{1}{4} \bigg(p(\sqrt{v_{i_0}} - 2\sqrt{v_{i_1}} < w < \sqrt{v_{i_0}})\\+p(w > \sqrt{v_{i_0}}  - \sqrt{v_{i_1}})
+ p(w > \sqrt{v_{i_0}} + 2\sqrt{v_{i_1}}) \\+ p(\sqrt{v_{i_0}} < w < \sqrt{v_{i_0}} + \sqrt{v_{i_1}}) \bigg).
\end{split}
\end{equation}
Thus,
		\begin{equation}
		\begin{split}
		P_e(\gamma_{i_0},\gamma_{i_1})=\frac{1}{4} \left(2\mathrm{erfc}(\sqrt{\gamma_{i_0}})+ \mathrm{erfc}(\sqrt{\gamma_{i_1}}-\sqrt{\gamma_{i_0}})\right.\\
		\left.-\mathrm{erfc}(2\sqrt{\gamma_{i_1}}-\sqrt{\gamma_{i_0}} )-\mathrm{erfc}(\sqrt{\gamma_{i_1}}+\sqrt{\gamma_{i_0}}  )\right.\\
		\left.+\mathrm{erfc}(2\sqrt{\gamma_{i_1}}+\sqrt{\gamma_{i_0}} )  \right).
		\end{split}
		\end{equation}
		
		\item Case 3: We assume that $\gamma_{i_0}=\gamma_{i_1}$. By using Figure 1.c, we obtain		
		\begin{equation}
		P_e(\gamma_{i_0},\gamma_{i_1})=\frac{1}{4} \left( \mathrm{erfc}(2\sqrt{\gamma_{i_0}})+1 \right).
		\end{equation}	
	\end{itemize}
	The BLEP is given by 
	\begin{equation}
	P_{E}(\gamma_{i_0},\gamma_{i_1})=1-\left( 1-P_e(\gamma_{i_0},\gamma_{i_1}) \right)^L,
	\end{equation}
	where $L$ is the number of bits per block.
	Notice that we can also consider the optimization problem which minimizes the overall average BEP by replacing in (\ref{PET_K1}), $P_{E}(\gamma_{i_0})$ by $P_{e}(\gamma_{i_0})$ and $P_{E}(\gamma_{i_0},\gamma_{i_1})$ by $P_{e}(\gamma_{i_0},\gamma_{i_1})$.
	Notice also that when the multiuser detector handles more that three packets in collision ($K \geq 2$), the analytical BEP expressions can be derived when the conditions of this case study are considered by following the same reasoning. We do not give theses expressions here for the sake of simplicity. 
	\begin{figure}[tb]
		\centerline{\includegraphics[height=7cm, width=7cm]{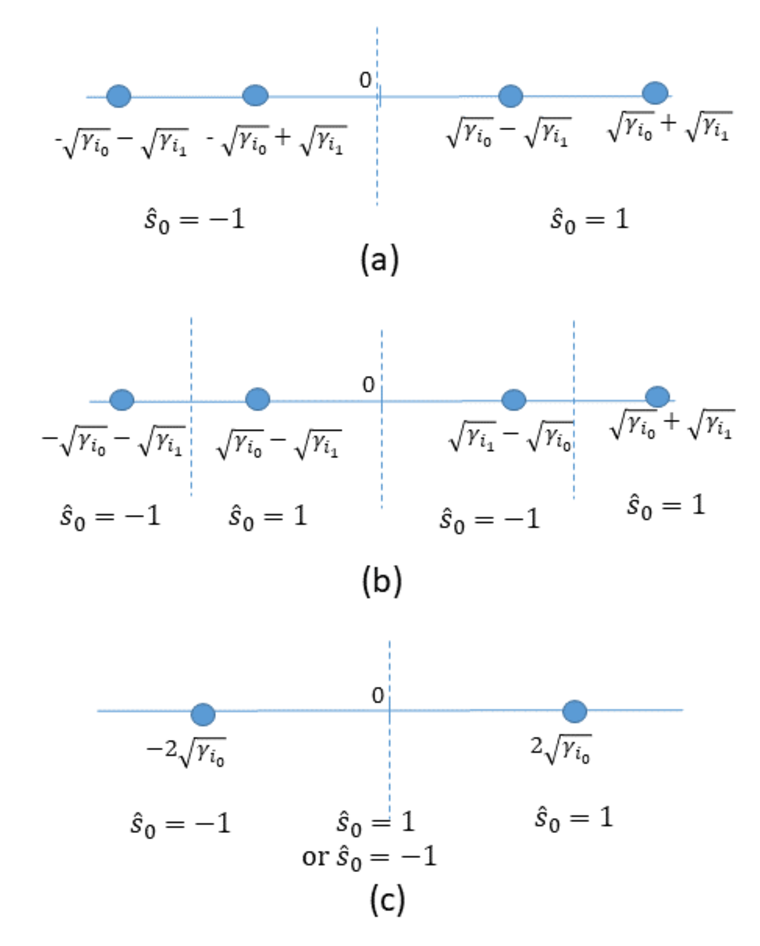}}
		\caption{Received constellation and decision boundaries. (a) $\gamma_{i_0}>\gamma_{i_1}$. (b) $\gamma_{i_0}<\gamma_{i_1}$. (c) $\gamma_{i_0}=\gamma_{i_1}$. }
		\label{fig1}
	\end{figure}
 	\section{Simulation results}
\begin{figure}[tb]
	\centerline{\includegraphics[width=7cm]{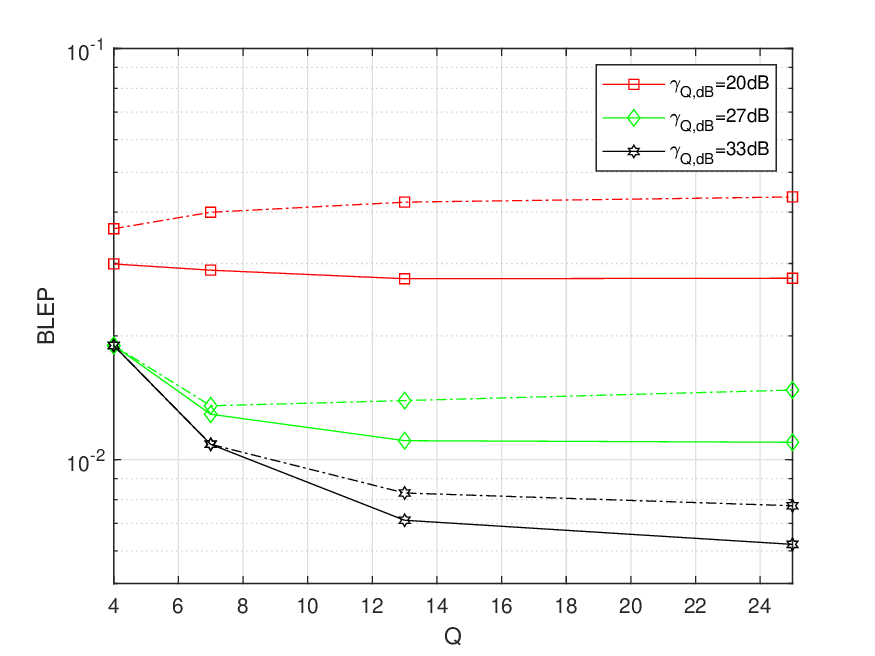}}
	\caption{BLEP versus $Q$ for BPSK modulation, different values of $\gamma_Q$ and $K=1$, when the optimized distribution is used (solid curves) and when the uniform distribution is used (dashed curves)    }
	\label{fig2}
\end{figure}

\begin{figure}[tb]
	\centerline{\includegraphics[width=7cm]{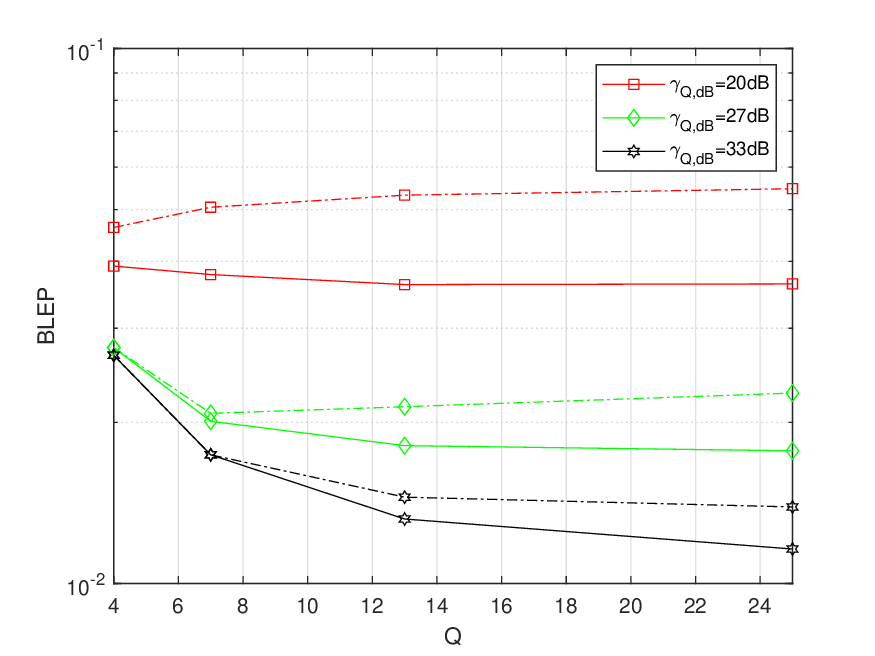}}
\caption{BLEP versus $Q$ for BPSK modulation, different values of $\gamma_Q$ and $K=2$, when the optimized distribution is used (solid curves) and when the uniform distribution is used (dashed curves)    }
	\label{fig3}
\end{figure}

\begin{figure}[tb]
	\centerline{\includegraphics[width=7cm]{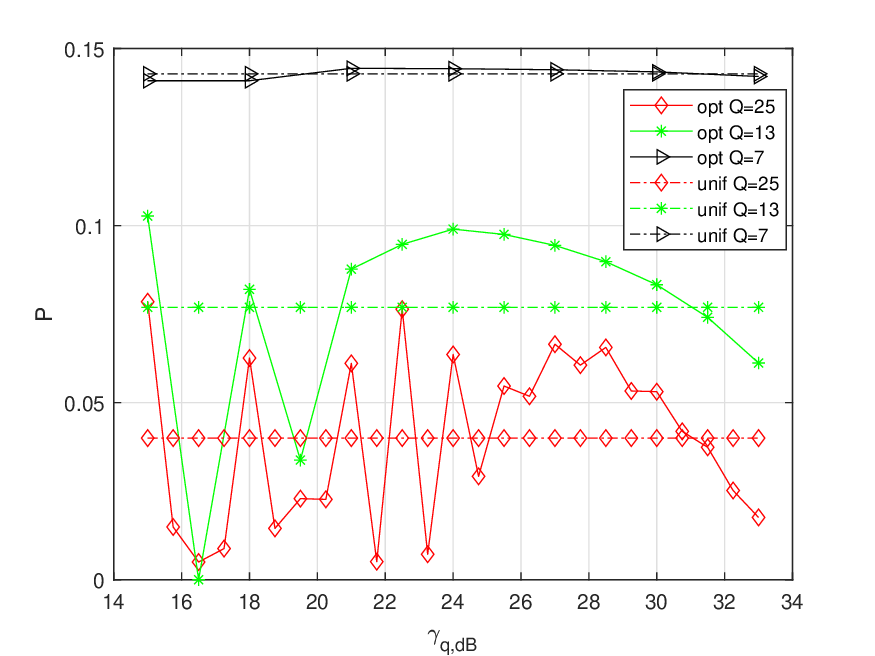}}
	\caption{Distribution of the received SNR at the BS for BPSK modulation, $K=1$, $\gamma_{Q,dB}=33dB$ and different values of $Q$: the optimized distribution (solid curves) and uniform distribution (dashed curves)   }
	\label{fig4}
\end{figure}

\begin{figure}[tb]
	\centerline{\includegraphics[width=7cm]{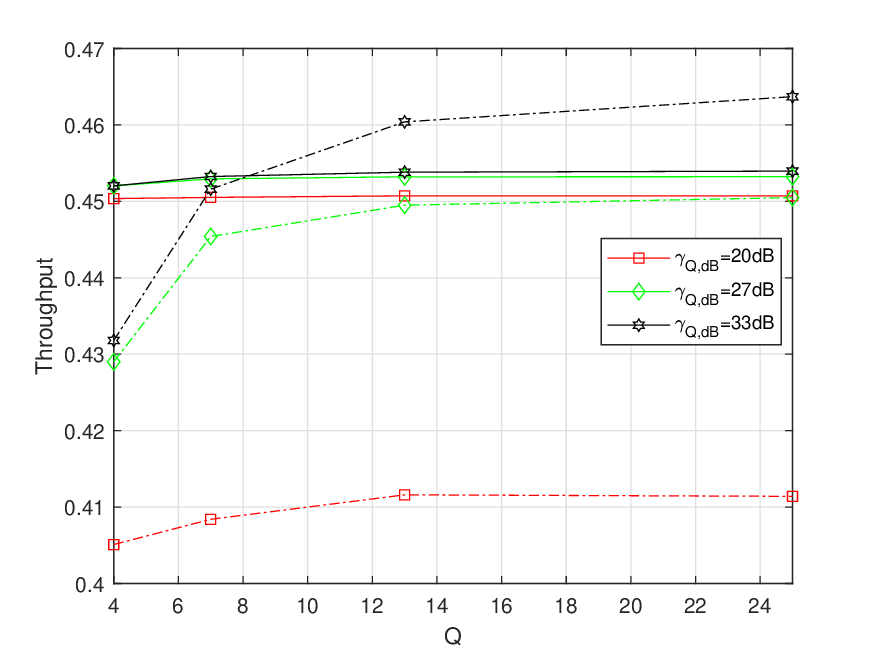}}
	\caption{Throughput versus $Q$ for BPSK modulation, different values of $\gamma_Q$ when the optimized distribution is used for $K=1$  (solid curves) and $K=2$ (dashed curves)   }
	\label{fig5}
\end{figure}

\begin{figure}[tb]
	\centerline{\includegraphics[width=7cm]{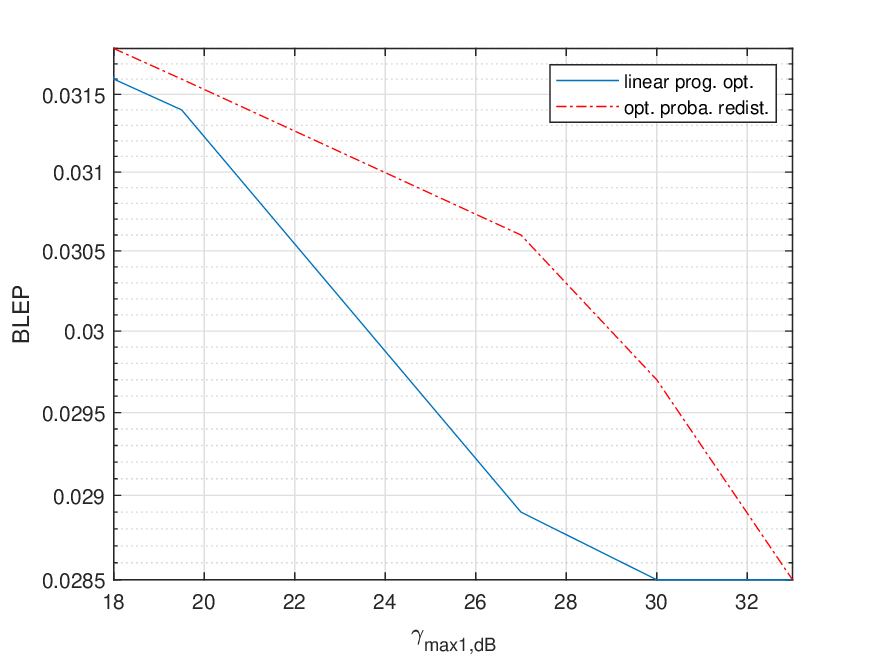}}
	\caption{BLEP versus $\gamma_{max1,dB}$, for BPSK modulation, $Q=13$ and $K=1$, when linear programming problem is considered (solid curves) and when the optimized probabilities are redistributed (dashed curves).}
	\label{fig6}
\end{figure}

\begin{figure}[tb]
	\centerline{\includegraphics[width=7cm]{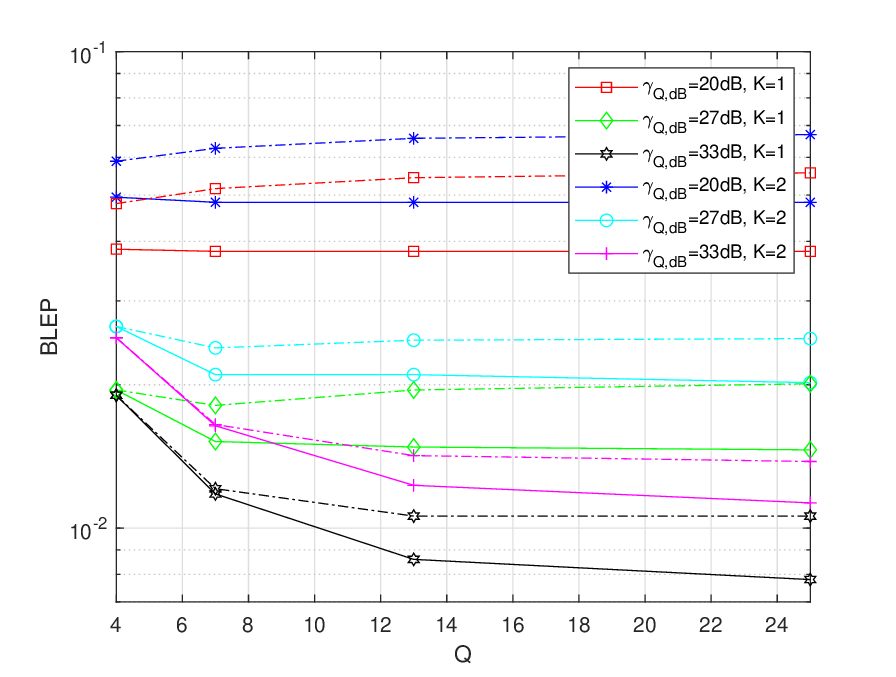}}
	\caption{BLEP versus $Q$ for QPSK modulation, different values of $\gamma_Q$ and $K\in \{1,2\}$, when the optimized distribution is used (solid curves) and when the uniform distribution is used (dashed curves)     }
	\label{fig7}
\end{figure}

 In our simulations, we consider the cases where $K \in \{1,2\}$, $\lambda=0.5$ and $L=100$. We assume that the possible received SNR levels are $\gamma_{q}=\frac{v_q}{N_0}$, for $1 \leq q \leq Q$, corresponding to the decibel values $\gamma_{q,dB}=10\log(\gamma_q)$. We set $\gamma_{1,dB}=15dB$. We choose, without loss of generality, the SNR levels as $\gamma_{q,dB}=\gamma_{q-1,dB}+\delta$, for $2 \leq q \leq Q$, where $\delta$ is a positive constant. We assume that $\overline{\gamma}$ is equal to the average value of $\underline{\gamma}$. We start by considering the assumptions used in the case study of section IV. First, we assume that there are no constraints on the users' transmitted powers. 
 	  Figures \ref{fig2} and \ref{fig3} show BLEP curves (obtained by using (\ref{PET})) versus $Q$, for different values of $\gamma_Q$, when the optimized distribution is used (solid curves) and when the uniform distribution considered in \cite{Choi2017} is used with $P_q=\frac{1}{Q}$ for $1\leq q \leq Q$ (dashed curves), for $K=1$ and $K=2$ respectively. We notice that when $Q\leq 7$ and $\gamma_{Q,dB} \in \{27dB,33dB\}$, both distributions lead to almost the same performance. This behavior occurs because, at low values of $Q$, the degrees of freedom available for optimization are limited, resulting in an optimized distribution which is roughly similar to the uniform distribution, as will be seen in Figure \ref{fig4}. However, when $Q>7$, we notice that our solution achieves a significant performance gain compared to the classical solution proposed in \cite{Choi2017}. Increasing $\gamma_Q$ improves the BLEP performance since it allows higher transmission power levels.
       We also remark that when $\gamma_{Q,dB}=20dB$ and then the range of SNR levels is narrow, the BLEP increases as $Q$ increases when the uniform distribution is used. This is due to the fact that as $Q$ increases, the spacing between consecutive SNR levels $\gamma_{q}$ and $\gamma_{q+1}$, for $1 \leq q \leq Q-1$, decreases. Then, $P_e(\gamma_{q},\gamma_{q+1})$ increases, particularly for low values of $\gamma_{q}$, since it becomes more challenging for the multiuser detector to separate colliding users' packets. The same reasoning holds when $\gamma_{Q,dB}=27dB$ and $Q>7$. In contrast, increasing $Q$ improves the BLEP when our solution is used, since the optimization mitigates the impact of narrower SNR spacing by strategically allocating probabilities to different levels. Figure \ref{fig4} shows the optimized distribution of the received SNR at the BS (solid curves) and the uniform one (dashed curves),  for $K=1$, $\gamma_{Q,dB}=33dB$ and different values of $Q$. We notice that when $Q$ is low ($Q=7$), both distributions are roughly equal which explains that in this case, both solutions achieve almost the same performance.
 	 Figure \ref{fig5} shows the throughput curves versus $Q$, when the optimized distribution is used, for $K=1$ (solid curves) and $K=2$ (dashed curves), for $\gamma_{Q,dB} \in \{20dB,27dB,33dB\}$. We notice that when the multiuser detection algorithm handles 3 users' packets in collision ($K=2$), the throughput improves compared to the case where $K=1$ when $\gamma_Q=33dB$ and $Q \geq 8$.    
 	 In Figure \ref{fig6}, we consider the practical case where there are constraints on the users' powers. We set $K=1$ and $\gamma_{Q,dB}=33dB$. We assume that the constraints do not affect the minimal SNR for both users, namely user 0 and user 1, as well as the maximal received SNR for user 0. We consider the case where user 1 can be far from the BS and then it can not use high power values due to the limitation on its maximum transmitted power (see the Remark). We assume that $p_{max}$ is set such as the maximum value of $\gamma_q$ is $\gamma_{max1}$ for user 1. Figure \ref{fig6} shows the BLEP curves versus $\gamma_{max1,dB}$ when linear programming problem is considered (solid curve) and when the optimized probabilities are redistributed among allowed power levels (dashed curve). We notice that as $\gamma_{max1}$ increases, the BLEP decreases and that considering a new optimization problem (linear programming) leads to a slight performance improvement.
 Figure \ref{fig7} shows BLEP curves versus $Q$, when the quadrature phase shift keying (QPSK) modulation is considered, for different values of $\gamma_Q$, when the optimized and uniform distributions are used, for $K\in \{1,2\}$. The obtained curves exhibit the same behavior as those obtained when BPSK modulation is used, confirming the efficiency of our method.
 	
 		\section{Conclusion}
 		In this letter, we have proposed to optimize the set of probabilities of selection of power levels in uncoordinated uplink NOMA systems in order to minimize the BLEP. Our solution is simple and can be applied for any multiuser detection algorithm and any set of power levels. Simulation results show that our proposed scheme using optimized distribution leads to a significant performance gain compared to the classical one using uniform distribution.




\begin{thebibliography}{00}
\bibitem{Liu2017}	Y. Liu, Z. Qin, M. Elkashlan, Z. Ding, A. Nallanathan and L. Hanzo, "Nonorthogonal Multiple Access for 5G and Beyond," \textit{ Proceedings of the IEEE}, vol. 105, no. 12, pp. 2347-2381, Dec. 2017, doi: 10.1109/JPROC.2017.2768666.
\bibitem{ChoiISWCS2017}	J. Choi, "NOMA: Principles and recent results," \textit{International Symposium on Wireless Communication Systems (ISWCS)}, Bologna, Italy, 2017, pp. 349-354, doi: 10.1109/ISWCS.2017.8108138.

\bibitem{Mostafa2019}A. E. Mostafa, Y. Zhou and V. W. S. Wong, "Connection Density Maximization of Narrowband IoT Systems With NOMA," \textit{ IEEE Transactions on Wireless Communications}, vol. 18, no. 10, pp. 4708-4722, Oct. 2019, doi: 10.1109/TWC.2019.2927666.
\bibitem{Tegos2020} S. A. Tegos, P. D. Diamantoulakis, A. S. Lioumpas, P. G. Sarigiannidis and G. K. Karagiannidis, "Slotted ALOHA With NOMA for the Next Generation IoT," \textit{IEEE Transactions on Communications}, vol. 68, no. 10, pp. 6289-6301, Oct. 2020.

\bibitem{Choi2017}	J. Choi, "NOMA-Based Random Access With Multichannel ALOHA," \textit{IEEE Journal on Selected Areas in Communications}, vol. 35, no. 12, pp. 2736-2743, Dec. 2017.
\bibitem{Wunder2015}G. Wunder, H. Boche, T. Strohmer and P. Jung, "Sparse Signal Processing Concepts for Efficient 5G System Design," \textit{ IEEE Access}, vol. 3, pp. 195-208, 2015, doi: 10.1109/ACCESS.2015.2407194.
\bibitem{Mai2021} L. Mai, Q. Zhang and J. Qin, "System Throughput Maximization of Uplink NOMA Random Access Systems," \textit{IEEE Communications Letters}, vol. 25, no. 11, pp. 3654-3658, Nov. 2021.
\bibitem{Semira2021} H. Semira, F. Kara, H. Kaya and H. Yanikomeroglu, "Multi-User Joint Maximum-Likelihood Detection in Uplink NOMA-IoT Networks: Removing the Error Floor," \textit{IEEE Wireless Communications Letters}, vol. 10, no. 11, pp. 2459-2463, Nov. 2021, doi: 10.1109/LWC.2021.3103937. 
\bibitem{Boyd2004} S. Boyd and L. Vandenberghe, Convex Optimization. Cambridge University Press, 2004.





\end{thebibliography}
\end{document}